\documentclass[11pt,twoside	]{article}
\usepackage{balance}  % to better equalize the last page
\usepackage{graphicx} % for EPS, load graphicx instead
\usepackage{url}      % llt: nicely formatted URLs
\usepackage[margin=1in]{geometry}
\usepackage{relsize}
\usepackage{wrapfig}
\usepackage{natbib}
\usepackage[normalem]{ulem}
\usepackage{tabularx}
	
\usepackage{url}      % llt: nicely formatted URLs
\usepackage{verbatim}
\usepackage[normalem]{ulem}
\usepackage{comment}

\usepackage{titlesec}

\usepackage{color}
\usepackage{array}
\newcommand{\textred}[1]{\textcolor{red}{#1}}
\ifx\noeditingmarks\undefined
   \newcommand{\pgwrapper}[2]{\textred{#1 #2}}
\else
   \newcommand{\pgwrapper}[2]{}
\fi

\usepackage{csquotes}
\usepackage[T1]{fontenc}
\usepackage[utf8]{inputenc}

\makeatletter
\def\url@leostyle{%
  \@ifundefined{selectfont}{\def\UrlFont{\sf}}{\def\UrlFont{\smaller \bf\ttfamily}}}
\makeatother
\def\pprw{8.5in}
\def\pprh{11in}

\usepackage[pdftex]{hyperref}
\hypersetup{
pdftitle={proceedingsform},
pdfauthor={LaTeX},
pdfkeywords={proc},
bookmarksnumbered,
pdfstartview={FitH},
colorlinks,
citecolor=black,
filecolor=black,
linkcolor=black,
urlcolor=black,
breaklinks=true,
}

\title{\textbf{Of Two Minds, Multiple Addresses, and One History: Characterizing Opinions, Knowledge, and Perceptions of Bitcoin Across Groups}}
\author{Xianyi Gao, Gradeigh D. Clark, Janne Lindqvist \\ Rutgers University}
\date{}

\begin{document}

\maketitle

\begin{abstract}
Digital currencies represent a new method for exchange and investment that differs strongly from any other fiat money seen throughout history.  A digital currency makes it  possible to perform all financial transactions without the intervention of a third party to act as an arbiter of verification; payments can be made between two people with degrees of anonymity, across continents, at any denomination, and without any transaction fees going to a central authority. The most successful example of this is Bitcoin, introduced in 2008, which has experienced a recent boom of popularity, media attention, and investment. With this surge of attention, we became interested in finding out how people both inside and outside the Bitcoin community perceive Bitcoin -- what do they think of it, how do they feel, and how knowledgeable they are. Towards this end, we conducted the first interview study (N = 20) with participants to discuss Bitcoin and other related financial topics. Some of our major findings include: not understanding how Bitcoin works is not a barrier for entry, although non-user participants claim it would be for them and that user participants are in a state of cognitive dissonance concerning the role of governments in the system.  Our findings, overall, contribute to knowledge concerning Bitcoin and attitudes towards digital currencies in general.

\end{abstract}

\section{Introduction}
Payment methods have been on a steady march towards focusing on user convenience since the advent of electronic and online banking reduced the overall friction in managing one's personal finances. As such, today has seen the decline of old payment methods like cash, checks, and money orders. This decline has given rise to new ways of managing payments in an interconnected, mobile world. Along with this shift towards online banking has come the next generation of money: digital currencies\footnote{also known as virtual currencies and cryptocurrencies.}, the most successful of which (to date) is Bitcoin.

Bitcoin~\cite{satoshi} is a decentralized digital currency proposed in 2008 that relies on the collective work of a distributed peer-to-peer network to maintain a public ledger of account history for all of the participants and to verify transactions amongst the different agents. Bitcoin is interesting as a digital currency as it represents not only a new way of managing finances but a new type of money that is different from all the fiat currencies seen in history. With digital currency, it is possible to perform all financial transactions without the intervention of a third party to act as an arbiter of verification. Payments can be made between two people with degrees of anonymity, across continents, at any denomination, and without any transaction fees going to a central authority (e.g.~a bank). However, it only recently started garnering mainstream attention from news outlets and financial institutions alike for its promise and volatility.

The upper bound estimate on the number of Bitcoin users is about 1.8 million~\cite{users} with at least 12.8 million coins in circulation~\cite{coins} as of this writing. Evidently, the number of adopters is still low compared to long-standing fiat currencies. We became curious about what people both inside and outside the community think about Bitcoin and how they perceive it. Towards this end, we conducted a comprehensive semi-structured interview study of ten people recruited from various online Bitcoin communities (user participants) and ten people from outside the Bitcoin community around the United States (non-user participants) and surveyed them about a range of topics, including: opinions about Bitcoin, why they do or do not use Bitcoin, technical questions about the protocol, security and privacy concerns, government involvement, opinions about current payment systems,  and what aspects would they include in an ideal payment method.  Interestingly, we note that HCI researchers have not spent a lot of effort on studying the effects of money in applied purposes~\cite{kaye2}.

In this paper, we present the following contributions: (i) results from semi-structured interviews of ten non-users and ten users of Bitcoin about a variety of topics; (ii) non-user participants claim that they cannot use Bitcoin since they do not understand it, despite not understanding how other methods work in any great details; (iii) user participants do not understand the mechanics of Bitcoin very well;  (iv) user participant misconceptions about Bitcoin's security and privacy controls; (v) a preference for government insurance of deposits without any desire for regulation; (vi) a mapping between participant suggestions for an ideal payment system and Bitcoin.\\
\textbf{Note 1:} We use ``Bitcoin'' to refer to the system as a whole and ``bitcoin'' to refer to the unit of account in the system.\\
\textbf{Note 2:} We refer to participants recruited from Bitcoin communities for the express purpose of discussing Bitcoin as ``user participants'' (or ``users'') and participants recruited outside of Bitcoin commmunities as ``non-user participants'' (or ``non-users'').

\section{Bitcoin Fundamentals}
In this section, we give a high-level description of the Bitcoin protocol. Any term that is bolded is a specific technical term that is part of standard Bitcoin parlance.

Bitcoin is a peer-to-peer version of electronic cash that allows online payments to be sent directly from one party to another without going through any financial institution~\cite{satoshi}. Digital currencies have previously been plagued by the \textbf{double spending problem}, which means spending the same unit of money at two different points. Electronic representations of money that are directly analogous to physical currency dictate that the money must be represented as data on a drive; however, this data can be copied when transferring ownership and thus this becomes a problem of trust. When dealing with physical currency, trust is obtained by holding and visibly inspecting the bank note. Trust is achieved with electronic payment methods (credit, debit) through use of a third party (e.g.~a bank) to confirm the movement of funds. Digital currencies solve this problem by distributing trust among every agent in the system.

Bitcoin distributes trust by having the majority of participants agree on a global, public ledger of accounts that contains all movement of money between people since the system's inception -- this ledger is referred to as the \textbf{blockchain} (see below). A person determines the amount of money they have by adding up all of their debits and credits from the public ledger; there is no file or object to store on their computer.  Agreeing on a global, public history is not enough -- a large set of dishonest people could agree on incorrect histories to control the system. Bitcoin prevents this from happening by requiring people to solve a computational problem in order to write to the blockchain.  

The computational problem involves finding an approximate pre-image of a hash\footnote{Bitcoin uses SHA-256 for all hashing purposes.} whose hex value is less than some constant set by the system. The process for finding the hash involves assembling a \textbf{block}, which contains a nonce, a collection of transactions (see below), and various metadata\footnote{The metadata includes: a timestamp, a hash of the previous block, the Merkle root hash of previous blocks, the difficulty number, and the version number.}, and repeatedly hashing and modifying the block until the hash hex value is less than the system-defined constant. The blockchain is so-named because what is written to the ledger are these blocks; the history is one long daisy-chain of them.

When a person wants to transfer money, they broadcast a message to the network stating their intention -- this message is called a \textbf{transaction}. The simplest representation of a transaction consists of three components: an amount of bitcoins to send, an \textbf{address} representing a person to whom bitcoins are being sent, and a private key used to sign and verify the transaction. The \textbf{address} is a public key owned by a recipient -- the recipient uses their private key to send the bitcoins at that address off somewhere else.  Addresses are generally used only once in Bitcoin, and as such users will keep a collection of private keys for redeeming bitcoins at those addresses in a \textbf{wallet}\footnote{A wallet will also contain a variety of other information: all transactions between addresses, a pre-generated queue of public-private keypairs for future transactions, etc.}.

The people who try to solve the computational problem are called \textbf{miners} and the process of finding the solution itself is referred to as \textbf{mining}. This process of searching for block solutions is computationally wasteful as miners incur losses in the form of electricity, heating, and bandwidth. As such, miners are rewarded with a set number of bitcoins to incentivize them to verify the transactions. This additionally has the dual purpose of introducing bitcoins into circulation.

The total supply of bitcoins is designed to be kept constant at 21 million. The system is tuned, based on the performance of the miners, to create six blocks per hour. Since the system is meant to analogous to gold, which has a fixed supply and would be more difficult to find as more of it is mined, the number of awarded bitcoins per block halves after certain block goalposts are met\footnote{As of this writing, the block reward is 25 bitcoins; the reward will fall to 12.50 bitcoins once 367,500 blocks have been mined.}.

Bitcoin claims to have near-zero transaction fees since there is no  centralized third party (e.g.~a bank) imposing fees for moving money between accounts. Furthermore, privacy is meant to be given to all participants through removing all identifying markers and just representing users by transactions and addresses on the public ledger. Neither of these contain metadata that \emph{alone} can be used to deanonymize participants. However, transaction fees do exist in this structure -- people can pay a premium to have their transactions verified more quickly by aggressive miners. There is also publicly scrapable metadata on the internet as well as network traffic that can be monitored in order to deanonymize people through the blockchain (please see Background and Related Work below).

\section{Background and Related Work}

In this section, we discuss both background and related work on Bitcoin,  modifications of the protocol, and prior studies on payment systems.

\subsection{Bitcoin}

There has been myriad work examining the economics of Bitcoin. 
Bitcoin, despite being intangible, has economic value: it remedies the double spending problem, has low transaction fees, and detects fraud through public authentication~\cite{VanAlstyne:2014:WBV:2594413.2594288}.  The price formation model of Bitcoin appears to follow standard supply-and-demand, though demand is the larger determinant given the supply is limited~\cite{ciaian2014economics}. Bitcoin's market impact model appears to fit well with statistical latent order-book models~\cite{donier2014million}.  It remains unclear if consumers will treat Bitcoin as a currency replacement or as a commodity for speculation in the long-term~\cite{Cusumano:2014:BE:2661061.2661047}. Other work has analyzed: Bitcoin's money flow and wealth accumulation~\cite{kondor2014rich}, using Bitcoin as a measurement of socio-economic signals~\cite{garcia2014digital}, Bitcoin's economic limitations as a decentralized currency~\cite{cryptoeprint:2013:829, evans2014economic}, and nowcasting the close-price with social media chatter~\cite{kaminski2014nowcasting}. 

Decentralization and a lack of regulation is a core design principle of the Bitcoin protocol, though it is starting to come under debate. One attempt at inserting a regulatory organ into Bitcoin involves redlisting wallets linked to criminal transactions and giving miners the option to approve transactions to and from the redlist~\cite{pouwelse2014operational}. Several countries (USA, Germany, Sweden, etc.)~have started initiating their own efforts to regulate and tax Bitcoin profits~\cite{krohn2013practical}. 

It is claimed that the large computing power needed to control the public ledger is the chief deterrent towards colluding, dishonest mining strategies, and other attacks. This assertion has since been challenged, and it has been demonstrated that a colluding pool with a minority share of the computing power can manipulate the system~\cite{eyal}. Minority collusion is done by controlling information about block solutions; there is no incentive for nodes to broadcast block solutions if nodes wish to hoard transaction fees~\cite{babaioff}. In addition, large mining pools that are open to anyone may attack each other to increase their own revenue; this may result in diminishing open pools' revenue as a whole if the balance breaks~\cite{eyal2014miner}.
Large pools are also more likely to be targeted by distributed denial-of-service (DDoS) attacks~\cite{johnson2014game}, which can have serious effects on the Bitcoin economy~\cite{vasek2014empirical}.

Bitcoin has also drawn the attention of fraudsters; a study has shown that out of 192 scams, 20\% were associated with Bitcoin addresses and 13,0000 victims have been swindled out of at least \$11 million~\cite{Moore12}. Bitcoin exchanges are highly susceptible to attacks (currently 18 have completely closed down), and popular exchanges are more likely to suffer a security breach~\cite{moore2013beware}.

The Bitcoin system purportedly preserves anonymity by only identifying users by transactions between addresses that hold bitcoins without any other personal information~\cite{Hobson:2013:BIT:2517249.2510124}. However, it has been shown that wallet addresses can be combined with publicly scrapable metadata from Bitcoin discussion sites to identify users~\cite{spagnuolo}. Notably, the authors of this (called BitIodine~\cite{spagnuolo}) have potentially identified the wallet addresses of the Dread Pirate Roberts (DPR)~\cite{dpr} from Silk Road~\cite{silk}, hitmen hired by DPR, as well as addresses used by CryptoLocker ransomers~\cite{cryptolocker}.  A heuristic clustering based on changing wallets can also be used to de-anonymize people~\cite{Meiklejohn:2013:FBC:2504730.2504747}. By monitoring the traffic of the network it is possible to map IP addresses to transactions~\cite{koshy}.

There are other currencies influenced by the Bitcoin protocol to give additional privacy and security. FawkesCoin~\cite{bonneau} opts to use the Guy Fawkes signature protocol~\cite{anderson} to reduce dependency on elliptic curve cryptography. MixCoin~\cite{bonneau2} tries to address privacy concerns of users and the tracking-of-transactions~\cite{spagnuolo} by mixing addresses through a network prior to transactions such that they appear to be coming from a random address every time.

A recent study examining the usability of the key management required for Bitcoin across six different clients concluded that there has been innovation in usable key management with respect to Bitcoin but more work is still needed~\cite{Eskandari2015}. A comprehensive survey and analysis of current academic research into Bitcoin resulted in an outline for future threads of research~\cite{BonneauNew}. Revealingly, this work has shown that there has not been studies on motivations and attitudes of Bitcoin users and non-users.

\subsection{Payment Studies}

Payment studies have focused on how members of populations use payment systems. Populations of advanced age (eighty years or older) have shown that people mistrust newer or unknown methods of payments~\cite{vines} (online banking is the case in this study). In one way to address this issue of mistrust, aging populations were presented with a system called Cheque Mates~\cite{vines2}, wherein a digital pen would transfer handwritten checks to a computer to be verified via crowdsourcing.

General lessons on digital currencies can be gleaned from studying cultures with an abundant amount of them. Based on a study of Japan, which has many digital forms of money~\cite{mainwaring}, digital money should: reduce commotion, be centered around public use, support money management without increasing burden or degrading user experience, engage multiple senses, and be fun to use~\cite{mainwaring}. 

Studies have shown that emotional and historical experiences play just as important a role as economic self-interest when people make financial decisions~\cite{kaye}.
Finally, there have been studies about monetary transactions through the lens of social context: how money is used in poorer areas~\cite{collins2009portfolios}, how people make decisions to allocate their funds~\cite{zelizer1997social}, how marriage affects financial status~\cite{stocks2007modern}, how the elderly perform modern banking~\cite{vines2011eighty}, and how to optimize user interfaces for newly adopted mobile payment schemes~\cite{export143165}.
\section{Method}
In this section, we describe our recruitment process, the demographics of our participants, our interview procedure, and the interview coding.

\subsection{Participants}

We recruited 20 participants aged 18 years or older across the United States, of which 10 were Bitcoin users and 10 were non-users. Bitcoin users were recruited online from Bitcointalk~\cite{bitcointalk} and Reddit~\cite{reddit} and non-users were recruited on our university campus via flyers and online using Craigslist~\cite{craigslist}. We created two different sets of recruiting material for our two groups: (i) our recruitment material for Bitcoin users explicitly stated our interest in interviewing people who actively used Bitcoin, and (ii) our recruitment material for non-users participants stated that we were conducting a study on currency and payment systems and were interested in interviewing people to discuss these topics.

We refer to our recruited Bitcoin users as U1 to U10 and non-users as N1 to N10 (see Table~\ref{tab:demographic}). Our ten Bitcoin users were all male with seven from North America, two from Asia, and one from the Middle East.  From ten non-users, there were four females and six males, and 
seven of them were North Americans, one was Asian, and two were Indian.
Our users educational background varied from high school to college; our non-users' background varied from high school to graduate school.

\begin{table}[t!]
\small
\centering
\renewcommand{\arraystretch}{1.3}
\begin{tabular}{| m{1.5cm}  m{0.4cm}  m{0.4cm}  m{2.2cm} | m{1.5cm} m{0.4cm} m{0.4cm} m{2.2cm} |}
\noalign{\hrule height 1pt}
User & Sex & Age &  Occupation & Non-user & Sex & Age & Occupation\\
\hline
U1 & M & 44 & Engineer & N1 & M & 41 & Server\\

U2 & M & 21 & Student & N2 & F & 22 & Unemployed\\

U3 & M & 27 & Manager & N3 & M & 62 & Invent. Asst.\\

U4 & M & 41 & Marketer & N4 & F & 32 & Admin. Supp.\\

U5 & M & 35 & Manager & N5 & M & 37 & Admin. Asst.\\

U6 & M & 24 & Developer & N6 & F & 25 & Officer\\

U7 & M & 34 & Physician & N7 & M & 33 & Student\\

U8 & M & 29 & Self-employed & N8 & M & 37 & Counselor\\

U9 & M & 28 & Consultant & N9 & M & 23 & Technician\\

U10 & M & 22 & Self-employed & N10 & F & 24 & Student\\
\noalign{\hrule height 1pt}
\end{tabular}
\caption{Demographic information about our participants. U1-U10 refers to participants recruited from Bitcoin communities while N1-N10 refers to participants collected from outside of Bitcoin communities.
}
\label{tab:demographic}
\end{table}

\subsection{Interview Procedure}
Since our participants were spread across the United States, we opted to conduct semi-structured interviews in person, by phone, and through Skype. 
The interview questions focused on a participant's preference concerning  payment methods, their knowledge and opinions on digital currencies, awareness of Bitcoin, and any experience with Bitcoin transactions. If necessary, we also asked follow up questions to clarify the participants' responses. Each interview lasted for about 30 minutes with the audio for the interviews being recorded and transcribed to text.
Each participant was compensated with a \$10 VISA gift card for completing the study.

\subsection{Interview Coding}
After the interviews, we reviewed the audio transcriptions. Before annotating and categorizing, all the responses were read to familiarize the researchers with the range of ideas. Then, we annotated them for patterns. After several iterations of examination and assessment, we extracted key themes and associated original quotes to each theme. Some of the representative quotes will be shown in the Results section as we present the themes. At the end, we highlight the important features from the participants' responses. This approach helps to ensure reliability and to avoid biases in the findings.

\section{Findings}
In this section, we present the findings that emerged from our semi-structured interviews with our 20 participants

\subsection{Bitcoin Background}

We asked all of our participants, if they asserted that they knew what a bitcoin is, to define the term and explain briefly how the protocol works. About half (4) of our user participants answered using generalities; only one user participant spoke in detail.

The responses of five user participants reflected a partial understanding about Bitcoin: U1 said it is an online digital currency that has a fixed supply and is impossible to counterfeit; U5 framed his response using circular logic, saying: \textit{``Bitcoin is an address. ... It's an address that holds a certain amount of units of bitcoins.''}; U1 and U6 both defined Bitcoin as an Internet currency that is scarce, impossible to counterfeit, and has economic functionality similar to that of gold, with U1 saying: \textit{``Gold is scarce, and Bitcoin is also scarce, but if you wanted to pay a dollar in gold, you'd have to break up a piece of gold bar and hand it to someone. With Bitcoin, if you want to send me a dollar's worth of Bitcoin, it'd be as easy as sending an email.''}; U7 only said that Bitcoin needed a private cryptographic key to sign transactions; U10 knew that Bitcoin allowed people to send money globally without an intermediary, adding,  \textit{''It [Bitcoin] is using cryptography to secure itself.''}

Only one of our ten user participants phrased their response in a way that demonstrated deeper understanding.
 U9's explanation said that, first, an account would be needed from Bitcoin network to generate transactions for miners to verify. He explained the chain-of-custody: \textit{``you can also verify that the person created a Bitcoin according to the Bitcoin protocol. And therefore there's this consensus that this Bitcoin is legitimate and it was sent to you down this path of verifiable transactions. And now you, the holder of this private key on this Bitcoin, can then spend it and use it to make transactions on the Bitcoin network.''}

The remaining four users had misconceptions: U2 and U8 misrepresented how a bitcoin is defined, with U2 saying:  \textit{``I think it [a bitcoin] could be regarded as just something that sits like a piece of code in basically an electronic form. And it's considered to have a value which makes it a currency.''}; U3 and U4 both mentioned that Bitcoin is about solving a mathematical problem to get rewards, with U3 saying: \textit{``To the best of my knowledge, Bitcoin is a mathematical problem that has been solved by computer and the Bitcoin was the reward for solving the equation.''}.

A bitcoin is not computer code, but rather just a unit of account on the public ledger while the purpose of the protocol is not about solving mathematical problems for rewards -- that exists only to incentivize transaction verification.

Six non-user participants asserted that they had at least heard about Bitcoin;  the remaining four responded in the negative. N2 believed it was a currency used expressly for black market purchases: \textit{``I've heard of it. It is part [of] currency used for black market sales.''}; 
N3 knew of bitcoins through the context of a financial instrument, saying: \textit{``It's a new currency, basically. To me it seems very speculative. I personally would not want to be involved. Awful lot of risks.''}; N4 was aware that Bitcoin was growing in popularity, but did not know what it is; N9 and N7 referred to Bitcoin as a virtual currency, with N7 believing that it is ``technically complicated'' and ``difficult to use in practice'' without explaining why; N8 demonstrated the deepest awareness, stating: \textit{``It's a purely electronic currency. An individual Bitcoin is some ridiculously long number that's arrived at through a particular algorithm. I know people who do mining and they have their machines spending all of their extra cycles or even just all their cycles computing out these algorithms looking to find another end point that is a `bitcoin'.''}

\subsection{Bitcoin Usage}

Several themes emerged about how participants used Bitcoin.

\textbf{Patterns of Use:} We probed our user participants about how frequently they made transactions. Five of our participants (U1, U4, U6, U8, U9) did it heavily, two (U3, U7) used it occasionally, and three (U5, U2, 10) rarely used it. We define heavy use as daily, occasional use as at least twice a month, and infrequent as six months or longer.
Most users (besides U2 and U10) thought that transactions were easy to manage and perform.

We asked our non-users participants if they had ever considered trying to use bitcoins. Eight out of the ten said they did not know enough about Bitcoin to comfortably use it; the remaining two (N4 and N8) either did not feel the need to use them or were misinformed about how they could be used -- for example, N4 said: \textit{``I didn't need to pay with that. I could pay with what I have now. Yeah I've never needed to use Bitcoin. I just haven't had the need to utilize it yet.''}; N8 thought bitcoins could only be obtained from mining and was not sure who accepted them, stating: \textit{``Well, I don't have any bitcoins and I haven't really had any computers that are powerful enough to really do any effective bitcoin mining... I'm not sure who actually accepts bitcoins as payment.''}

\textbf{Mining:} The majority of our users (seven out of ten) had not participated in any mining, citing the high hardware requirements (U2, U9) and an overall lack of benefit (U3, U4, U6, U10) as the reasons why. Three users (U5, U7, U8) had tried individual mining as well as through pools,\footnote{Mining pools are a group of miners combining efforts for a single block solution. This grouping reduces the amount of reward each miner gets but drastically reduces the variance when it comes to actually finding a block.} but they eventually decided to stop mining due to diminishing returns. U8 said, \textit{``I ...  put my processor to work on a pool and made practically nothing. I might have mined something back in 2011 when I was first taking a look.''}

\textbf{Bitcoin: Investment or Currency?}
The Bitcoin community is split between people who speculate and those who believe in its long term potential. Five of our participants (U1, U3, U4, U5, and U8) treated Bitcoin as both an investment and a currency, with U5 opining: \textit{``It's like an investment in the way that you can see the price rise over time. You can hold it. You can sell it. Just like any other investment, if you buy and hold, you have a chance of your investment increasing or decreasing. As a currency, it's accepted at Overstock and Amazon and all that stuff, and you can use it as a way to acquire goods.''}; U2, U10, and N4 said Bitcoin was more like an investment -- U10 observed, \textit{``I think at this point it's more of an investment. I think the value is far too volatile to be buying it and holding it as a currency. So any money that you do put in Bitcoin, it should only be money you're OK with losing at this point. So if you're using that philosophy, I think it should be considered an investment.''}; 
 U6, U7, U9, and N3 thought Bitcoin was more like a currency, U7 commented: \textit{``... It's got a long way to go before it can be a useful high volume ... profit center. Right now there are a lot of technical aspects, the most basic of which is block size limit.''}; U9 mostly saw Bitcoin as a currency, adding: \textit{``whereas a lot of people used it for investment by exchanging with other currencies.'' }

When asked about whether bitcoins have value as an investment, eight out of the ten users indicated reasons that they believed so: personal experience (U1, U3), the upward trend of the price (U5), expectation and confidence (U6, U8), long-term popularity (U9, U10), and smart people (e.g.~developers) behind controlling the price (U4). Two users thought of investing in bitcoins as a risky venture, with U2 saying:\textit{``Definitely a risky one. I believe it's risky because it's still like a new concept. It can either flop or it can go up depending on how much people want to use it.''}
Only one non-user participant (N4) provided thoughts about investment potential, commenting that bitcoins appeared as a good investment to her because acquaintances had profited from investing in them.

\subsection{Security and Privacy Issues}

All user participants responded in the affirmative when queried about whether Bitcoin is secure, with U3 declaring: \textit{``If done properly, I would say so, because of the confirmation aspects of it ... to be able to get five hundred confirmations that my money is legitimate and it is now your money and it is now in your possession.''} When comparing to bank security, U3 continued: \textit{``Bitcoin is better protected, because I have my Bitcoins offline so unless someone finds where they are hidden, I do not think that they could be taken from me.''}; U9 also thought that a Bitcoin wallet was more secure than a bank, saying: \textit{``In defense of a Bitcoin wallet, you control yourself. You have access over your private keys. You generate your private keys yourself. This for sure is more private and secure than other sorts of financial transactions. ... [For banks,] you're trusting a third-party with your money and your activities.''} U7 thought bitcoins were more secure than using credit cards, stating:  \textit{``The credit card is insecure. People can see the numbers. Because of the cryptography in bitcoin, [it is] very secure. It's more secure than many things.''}

Other participants hedged on the question, responding that bitcoins are only protected as long as account information is properly managed;  U4, U5, U8, and U10 said Bitcoin should be secure as long as users exercised caution with their keys with U4 commenting: \textit{``They are as safe as the individual regulating them.''}  U5 also mentioned that transactions are reversible only on the good faith of the recipient and U4 stated there are evil agents in the system trying to steal bitcoins. 
Non-user participants could not answer the question, citing their lack of knowledge; only N4 responded in the affirmative, citing  the activity of developers working on the technology as his reason for believing so.

We then asked our participants about whether privacy is preserved in the network. Most Bitcoin users (nine out of ten) thought there is good privacy protection, with U7 comparing it to credit cards: \textit{``The credit card by definition is inprivate. You're giving away private information to someone in hopes that they don't use it again. Bitcoin security is on a person-to-person level. In my hands its much more secure and private.''} U1, on the other hand, believed that other cryptocurrencies had better privacy protection (e.g.~Vericoin~\cite{vericoin}), saying: \textit{``it [Vericoin] has anonymity capabilities built in, which is better than Bitcoin.''} All non-user participants declined to answer the question.

\subsection{Technical Aspects of Bitcoin}
We also probed our user participants about the fine-grained details of the Bitcoin protocol; we did not ask any of these questions to our non-user participants.

\textbf{Bitcoin mining:} The user participants had varying degrees of mining knowledge: half of them gave accurate responses, one demonstrated a high-level understanding, and one user did not understand how mining worked at all.

Responses from U1, U2, U6, U7, and U9 were all correct,  pointing out that the goal of Bitcoin mining is to maintain, update, and verify the transactions. 
\begin{quote}
U1: \textit{``Bitcoin mining is the act of maintaining and updating the transaction ledger for the currency and in the cost of doing that, they are receiving a share of the new coins that are being mined, as well as transaction fees.''}

U6: \textit{``What the coin miners do is [to] confirm transactions across the network. They do a lot of complicated computations to verify that I have the money in my account and that it hasn't been double spent anywhere. They make the Bitcoin work, and they put trust in their system by verifying transactions.''}
\end{quote}

U9 gave a more detailed description about mining: 
\begin{quote}
U9: \it{``Bitcoin mining is the process of building the blockchain which is the ledger of all of the transactions
that happen on the Bitcoin network and it's the process of doing this in a decentralized fashion. ... They
[miners] are taking a block which is really just a group of transactions and data. ... Miners take all the
valid transactions they want to include, put this into a chunk of data, hash that data with SHA2, which
creates some number and then they check this number to see if it meets a number known as the
difficulty based on some Bitcoin protocol. The number gets lower and lower over time [as] more and
more miners are mining. ... They create a chunk of data with valid Bitcoin
transactions, they hash it beneath this number according to the protocol. ... [They] publish or broadcast this
block. ... All other miners can verify that all the transactions included in it are valid ... This becomes a new
piece of the Bitcoin blockchain -- the ledger -- and the miner who found this block gets awarded. ... So, basically, there's no party that can control the transactions going to the Bitcoin
blockchain unless they hack into long-term control of over 50\% of the hashing power. ...''}
\end{quote}

U8 and U10 were unsure about the mining process, only mentioning mining as an ``important transactional process.'' U3 and U4 thought the point of mining was to solve mathematical problems and to get rewards. U5 was not sure and said mining could be ``finding new addresses for a bitcoin''.

\textbf{Transactions:} We then asked participants to define transactions in terms of the Bitcoin parlance and to explain the verification process. U1 provided a brief explanation, saying that \textit{``a Bitcoin transaction is just an entry in the ledger, moving bitcoins from one account to another.''}; 
 U4 and U5 seemed to avoid the question, only stating that the Bitcoin network ``verified transactions and ensured no double spending.''; U6 and U8 simply said transactions are the action of bitcoins moving from ``one Bitcoin wallet to another''.

Participant U9 again provided the most comprehensive explanation out of all user participants.
\begin{quote}
U9: \textit{``To do this, you need to have a private key which corresponds to a certain public key or Bitcoin
address. ... There would be a public key and address of someone who previously have sent bitcoins to
you. ... The person who has this private key can create a message. ... I'm signing this
message with a private key. Then I'm sending these bitcoins to this address and you specify a Bitcoin
address in this message. You sign the entire message with a private key. Whatever you have that doesn't
go into the account becomes the miner key and you sign the transaction, and then you broadcast it from
your branch. ... Once the majority of the Bitcoin network sees this transaction, they are assumed valid.
They will include that transaction in their next block, if they're mining. ... It permanently goes to the
Bitcoin blockchain. And now, whoever has the private key to that address that you just sent bitcoins. ... Do the same process. That's basically it.''}
\end{quote}

The remaining four user participants were unsure and declined to answer the question.

\textbf{Block:} We further asked our recruited user participants to define what a block means in terms of the Bitcoin protocol and found that only two users (U7 and U9) could answer in sufficient detail. 
\begin{quote}
U9:\it{ ``A block is a computer-based structure that contains a hash of the previous block ... When a block is solved, it produces bitcoins and gives [them] to the winner or winners. They [use] a bunch of valid transactions that have occurred at that point to validate new transaction inputs... There is also a nonce within the block, as well as [a] Merkle root... A Merkle tree is where you have the hash value of every leaf of a tree such that in order to be certain about the completeness or the accuracy of all the descendants along one path [so] you don't need all the data on every single leaf.''}
\end{quote}

Five user participants had limited knowledge: U1 and U6 commented on the block production time, with U1 saying: \textit{``The blocks are created once every ten minutes. Inside each block is all the transactions that have happened within the past ten minutes.''};

U8 said that rewards are given for finding block solutions; both U4 and U10 knew that each block at least contained a hash of previous transactions. 

The remaining three user participants could not properly define blocks: U2 and U5 said they did not know enough about the protocol; U3 believed that a block is the same as the miner's attempt to find a block solution, that is, a mathematical problem to be solved.

\textbf{Blockchain:} We then asked the user participants to define the blockchain: six users (U3, U4, U5, U6, U8, U10) knew that the blockchain represents a history of all transactions, with 
U5 explaining: \textit{``I would assume that would be the public ledger that shows all the Bitcoin transactions and how many times transactions just confirmed. It's like the blockchain is a protocol that allows for public ledger showing transactions.''}; U1, U7, and U9 said the blockchain is the accumulated blocks linking together. U2, however,  posited that the  ``blockchain might be a string of code''. Overall, user participants seemed to be more familiar with the term ``blockchain'' and what it represents as opposed to a block.

\subsection{Regulation}

We were curious about gauging our participants' feelings about the government regulating the currency given the breadth of recent scandals concering exchanges (e.g.~Mt.~Gox).

Five Bitcoin users (U1, U2, U5, U8, U10) felt that the government should provide protection against fraud: U1 thought that \textit{``the government should establish safeguards against fraud and abuse of financial institutions dealing with bitcoins.''}; U5 said that \textit{``[the government] could help people gain trust [in] Bitcoin by providing [financial] insurance. People tend to misunderstand Bitcoin because of its use on black markets, but people can do bad things with any type of money too.''}

Two users (U3, U9) could not decide one way or the other since they thought there were both advantages and disadvantages. U7 had a clear view of what should be regulated: \textit{``It could be that the exchange point between Bitcoin and the regular currency should be regulated. As far as regulating Bitcoin itself, I think it's a fool's errand.''} He also believed that  \textit{``being able to criminalize certain actions taken with money is very useful for society.''}

Two Bitcoin users (U4, U6) said it should not be regulated at all; U4 explained: \textit{``The entire concept and process was developed with the intention of not having a third party especially the big brother, regulating.''} Only three non-users (N2, N3, N4) responded to this question, all of them believing it should not be regulated -- however, their responses were less about the core principles of Bitcoin and more about their biases towards the government.. For example, N4 said: \textit{``No, because I think they do a bad job of regulating our own money.''}; N3 said, \textit{``No. I just don't think the US government should be involved in regulating anything, pretty much. I think they should just stay out of people's business, basically.''}

Six user participants and two non-user participants said the government should stay neutral, with U7 commenting: \textit{``I think laissez-faire is probably the best. Right now it is not appropriate for the average consumer. There's too many subtle technical things, difficulties in security, too much get rich quick hopes out there for it to be widely adopted as is.''}; N2 elaborated: \textit{``Maybe there's an advantage, advantage for the economy, maybe there's disadvantages. I think they should be indifferent.''}

  Four Bicoin user participants and one non-user participant believed the government should take a role in advocating for Bitcoin: U6 said, \textit{``It would be cool if they encouraged adoption. I'd rather have them encourage it than discourage it. But either way, in the long run, it won't matter.''} No participant wanted government to strictly prohibit the use of Bitcoin; the remaining seven non-users had no comments about the subject.

\subsection{Current Payment Systems}
We were also interested in our participants' thoughts about common payment methods in the United States, such as:  debit cards, credit cards, checks, Paypal, MoneyGram, and Western Union, to contrast with Bitcoin.
All 20 participants were asked about these methods.

\subsubsection{Common Payment Methods}

We started off by asking participants about their most commonly used payment method. Not surprisingly, most participants said either credit cards (U1, U3, U7, U9, U10, N3, N4, N7, N9, N10) or debit cards (U2, U4, U5, U6, N2, N5, N6, N8). The main reasons for using credit cards were: convenience (U1, U7, U9, U10, N3, N4), improving credit scores (N7), delaying payments (N10), and getting reward points (U3, N9). N3 said, \textit{``Credit card. Convenience. Just the fact of not having to carry a lot of cash around with me basically.''} N7 was trying to improve their credit score, saying: \textit{``Just to improve my credit score right now. I don't have any major expenses.''} The main reasons for using debit cards were for convenience (U2, U5, U6, N2, N6) and for avoiding interest or fees (U4, N5, N8). U4 and U5 also mentioned that cards were accepted almost everywhere -- however, U5 said he would prefer to use bitcoins more often, complaining: \textit{``Most places don't accept Bitcoin, so debit card becomes the next most convenient thing.''}

The minority opinion split between cash and checks: N1 preferred using cash, saying, \textit{``Cash. Because if I use a credit card, sometimes I don't know how much is [left] on the balance, and sometimes I might overdraw my card.''};  10 used checks the most because he needed to deal with business payments.

We then asked participants their least used payment methods, with participants responding: money orders, checks, Paypal, Western Union, and MoneyGram; they cited reasons such as  \textit{``clumsy to use''} (money orders and checks), \textit{``never tried using''} (MoneyGram), \textit{``expensive wiring fees''} (Western Union), and \textit{``extra fees [being] charged''} (Paypal).

When we asked participants about the payment method they preferred not to use, three participants (U5, N6, N8) said credit cards in spite of the fact they used them frequently. U5 said, \textit{``It just seems so dangerous. There's so much information attached to it and so much identity theft going out with credit cards.''}; N6 disliked dealing with interest payments.

We asked our participants questions about the mechanics of credit cards to gauge their understanding. Most participants (18 out of 20) asserted that they understand how credit card processing works. Eight of them (U1, U2, U4, U10, N4, N6, N7, N10) couched their explanations in the context of making purchases, with U4 stating: \textit{``I swipe it on the terminal, it's the cashier registers how much I'm to be charged on my account. Then that goes on my account and I subsequently get billed for it at the end of the month.''} 
three participants (U6, U7, U9) focused on explaining how merchants get the money through credit cards, with U6 saying: \textit{``On the merchant side, they have a point of sale system. They get charged per swipe. They're responsible for returning that money if there's any charge backs.''}. Others explained how credit is determined: based on income levels (N3, N8) and timely payments (N5, N9, U5); the remaining discussed peripherals: interests applied when lending (U8), and payments linkable to checking accounts (N2).

When we probed participants with follow-up questions about how transactions are processed and secured without compromising privacy they could not answer. For instance, N8 said: \textit{``To be honest with you, I'm not entirely sure where it goes after swiping through terminals...''}; N9 was not sure about the security and privacy aspects of credit cards: \textit{``I don't know [how they design the card security]. From what I understand, there is a chip or bar code [which] is the main part of the card having my information.''}

Several themes emerged from our participants' responses about advantages of electronic payment methods versus non-electronics methods; these are summarized below.

\textbf{Convenience:} 17 of our participants (exempting U3, U7, and N6) mentioned the convenience of using electronic payment methods, mainly pointing out that credit and debit cards are easier to carry, more convenient, and simpler than cash or checks. U5 said, \textit{``pros of electronic versus physical [non-electronic] would be that it's more convenient to use an electronic [method] rather than turn out money in change.''} Similarly, U8 said, \textit{``With credit it's much more convenient. I don't have to carry a pocket full of change. I don't have to carry bills. I get to direct the actual payment.''}

\textbf{Records:} Five participants (U2, U6, U9, N4, N7) said that electronic payment methods usually had monthly  records generated to keep track of expenses. U9 said, \textit{``[Credit cards have] this transaction history that is automatically generated which is useful if you're trying to budget your expenses, see where your money is going, or for business expenses. I suppose you get with that with a check, or something like that, but it's not as detailed as with your credit card statement.''} 

\textbf{Online shopping:} Four participants (U3, U7, N1, N7) pointed out that electronic payment methods had exclusivity with respect to online shopping. U7 mentioned, \textit{``...  advantages are that it's hard to use cash over the Internet.''} N1 also said, \textit{``You can do a lot of online shopping with it [credit card]. And there's like a lot of discounts that goes with a lot of online shopping.''} 

\textbf{Security:} Five participants (U5, U10, N6, N7, N9, N10) said that electronic payment methods were more secure than non-electronic methods. For example, N10 mentioned that cash could be physically stolen or lost while electronic methods could be better protected. She continued, \textit{``the incidence of thefts can be reduced to a great extent if you're using credit and debit cards.''} U5 said electronic payments verified immediately when shopping online or in person, while checks could be ``counterfeit or bounced back''. U10 added that electronic methods have better customer protection mechanisms such as chargebacks and record tracking.

The following sections concern themes that emerged when we asked about the disadvantages of using electronic payment methods as compared to physical ones.

\textbf{Security:} Participants still felt these methods were insecure despite enumerating several customer protections.  U1 was concerned about passwords being hacked when using online payment methods. U2 thought that there were risks for using electronic payments when needing to transfer money to \textit{``a middle man''}. N1, U3, and U7 were concerned about online identity theft. U5, U8, N8, and N9 mentioned concerns about personal information being leaked when cards are lost or stolen. N10 professed that losing a card is worse than losing cash.

\textbf{Overspending:} U4, U6, and U9 mentioned that it was easy to overspend when using credit or debit cards.U4 said, \textit{``You are spending more than you can afford. You don't really feel that you're spending money because its digital. When I spend cash in my wallet I see it disappear.''} 
\textbf{High fees and interests:} U9, U10, N2 and N5 mentioned that there are high fees and interest rates associated with credit cards, especially if there is an outstanding balance. As a business owner, U10 said: \textit{``Well, if I'm accepting any of those electronic payment methods, I'm subjected to getting payments reversed on me. That's a huge disadvantage. There's typically higher fees associated with those payment methods as well. [These are] certainly the two major drawbacks.''}

\textbf{Credit Limits:} N3 and N6 said that they needed to remain aware of their current credit limit on the cards before making purchases. For example, N6 said, \textit{``You got to take time out to find out how much you got in your balance in your card. That's the inconvenience right there.'' } This is mainly because outstanding fees are charged if customers go over the credit limit.

\subsubsection{Bitcoin Comparisons}

Many of the user participants compared Bitcoin to other payment methods in attempts to motivate why they preferred to use it. 

\textbf{Transaction Benefits:} U1, U4, U5, U9, and U10 mentioned that Bitcoin has faster transaction speeds than most traditional payment methods (exempting cash for person-to-person transactions). U1 said, \textit{``It's very efficient. It's fairly fast, faster than traditional financial systems.''} U4 compared Bitcoin transactions with Western Union: \textit{``It's almost immediate, if I was transferring money to somebody in Holland. If I was going to Western Union, paper work [needs to] go to the office, wait for that, and then wait for that whole process to end on the other side. With Bitcoin, it takes a minute or two.''}

U4, U9, and U10 mentioned that Bitcoin had lower transaction fees: U4 said, \textit{``The fees. It costs a very, very small [amount], a couple tens of a percent for every transaction.''}; U9 said, \textit{``[Bitcoin has] low transaction fees since no processing needs to go through a third party.''}

U9 and U10 also said that Bitcoin is easy to access and manage: U10 said, \textit{``Well, I can use it basically wherever I have Internet access, whether that's my phone, my laptop, and my tablet. I can instantly send payments. People can instantly verify that I've made the payment.''}

\textbf{Security and Privacy Benefits:} U1, U3, U5, and U9 said Bitcoin is very secure compared to traditional payment methods. U9 mentioned that Bitcoin addresses has stronger encryption methods than credit card payments, with almost no ways to hack it if properly handled. U5 said that, because of the the high computational power involved, \textit{``they cannot counterfeit them [bitcoins].''} U1 also pointed out that Bitcoin is very secure as long as users exercise common sense about their personal information and their wallets.

U3, U4, U6, U7, and U9 said Bitcoin is almost anonymous. U3 said, \textit{``You can be almost anonymous if you choose to do it that way. It's safe, it's easy to travel with.''} U4 said, \textit{``Even though it's not completely anonymous, it is in a way that unless somebody is trying to do a reverse analysis on [the] blockchain and really investigate and follow the transactions down the line, anything you buy or sell is unknown.''} U9's also mentioned privacy: \textit{``... the fact that there are privacy implications, that can be important. Not so much for a regular person, perhaps, but there are people really value their privacy for one reason or another and Bitcoin makes it possible.''}

\textbf{Ubiquity:} U8 said that Bitcoin has the potential to be a global currency; U9 furthereed that point by suggesting it be used as an anchor in countries where the local currency is not stable: \textit{``One good example is Argentina where there's so much manipulation with the local currency and so much inflation. You take yourself out of that if you have the money that you're hoping to spend in Bitcoin.''}\\

\textbf{Disadvantages of Bitcoin:}
U1, U2, U5, U8, and U10 opined  that Bitcoin lacks mainstream adoption. U2 said, \textit{``Definitely the fact that not many places or companies are taking it as of now. Since it's relatively still a new concept being introduced to people in the world.''}

U3 and U4 said that Bitcoin has no structure for reversing payments. U3 said, \textit{``There are no chargebacks. It can be both an advantage and a disadvantage. Right now, [it is] really just a fact that it's so new [such that] there's not a lot that has been made available for it. This is not necessarily innovative but not fully released and insured. ...''}

U6 and U9 pointed out that the price of a bitcoin is not stable. U6 gave an illustrative example:\textit{``If the [Bitcoin] price goes up to five thousand dollars tomorrow, then the fifty dollar toaster I bought [using bitcoins] would be the elephant in the room. Like, I could have bought a TV instead of a toaster. [I] fear of missing out on a bitcoin price increase.''}

Some participants thought that initiating Bitcoin transactions takes more steps than other payment methods.  U6 said that \textit{``it takes an extra couple steps in the process to complete a Bitcoin transaction compared to doing on one step [for a credit card] that already has my card information stored in.''} U7 also pointed out that \textit{``most of the time, initiating any transaction is slower with Bitcoin simply because of the security measures involved.''} Similarly, U9 thought that using Bitcoin is not as straight forward as other common payment methods. He said, \textit{``You need to have quite a decent grasp of the computer and technology to be able to use this safely and effectively. They could be lost or stolen if you're not careful. ... Everyone knows how to swipe a credit card and knows how to click buttons on PayPal but may not know what to do with private keys or what private keys even are.''} However, he also believed that more people would know how to use it in the near future.

\subsection{Ideal Payment System}

We asked all participants about what qualities an ideal payment system should have; several themes emerged from these discussions --  though we are careful to note that these questions were asked prior to any questions about Bitcoin in order to avoid biasing participants.

\textbf{Faster transactions:} Four participants (U3, U5, U8, N9) pointed out that, in any case besides a person-to-person cash transaction, the current system is slow to verify and process transactions (especially for international wire transfers). 

\textbf{Multiple storage modalities:} U2 believed that currency should be paid from a smartphone instead of wallets. He said, \textit{``There's a lot of applications on your phone, or tablet, that basically hold money for you and it's definitely a lot more convenient to just pay with that versus holding a wallet.''} Similarly, N6 wanted a payment system that \textit{``has both physical and electronic [elements] inside it''.} U6 also thought that a payment system should be available to view, manage, and spend with an application in the phone. What these participants described are similar to mobile payment methods, that is, Apple Pay~\cite{applepay} and Google Wallet~\cite{googlewallet}.

\textbf{Increased security and privacy:} U7 and U10 said that a payment system should protect the user's privacy and personal information while guaranteeing anonymity. U4 and N8 said the transactions should be more secure and (in the words of N8): provide a \textit{``balance between convenience and security''}. N10 thought that it should deter theft, stating:  \textit{``[It should be a] currency which is not easily reusable by someone else, so the theft is controlled.''} U9 mentioned Bitcoin when answering this question:\textit{``I think the closest thing we have now to an ideal payment system or ideal currency [is] the Bitcoin form. Bitcoin has a lot of qualities that make it very well suited for a currency. There are technical aspects to how it works. It might be slightly different but it's a currency where the supply isn't controlled by any one party. ... I would basically create something very much like Bitcoin.''}

\textbf{Increased reliability:} N9 and N4 said a system should be more reliable and have reduced opportunities for making mistakes. N9 said, \textit{``... The system would have to be more reliable. As long as it's a trustable source, that would be something that I would look into.''} 
\textbf{Lower fees:} U6, U10, N2, and N5 thought there should be less fees. N2 said, \textit{``To me, is that they don't charge me too much extra money, including transaction fees.''} Also, U6 said, \textit{``It shouldn't have outrageous fees to move money from one point to another. I've been transferring money from my savings to my checking account; then, I receive a notification from that bank that I've been transferring too much money between accounts and if I continue to do so, they'll charge me a fee.''}

\section{Discussion}

Below, we discuss the major findings of our study:  1) non-user participants did not understand Bitcoin, but neither did most user participants; 2)  most user participants thought Bitcoin had good security and privacy controls despite evidence to the contrary; 3) participants highly disapproved of government regulation but still want  governments to insure deposits; 4) participants' opinions about attributes of an ideal payment system map directly to properties that Bitcoin has; and 5) Bitcoin has barriers to overcome that make it difficult to be used for mainstream adoption.

Out of our ten non-user participants, we found that four of them had never heard about Bitcoin while the remaining six were aware of it from a superficial level based on what they had heard from the media or their social network.

Three out of the six noted correctly that it is a type of virtual currency, two of them voiced concerns about the negative connotations it had in the media and wondered aloud about its legality.

Many of the non-users perceived Bitcoin as being technically complicated, hard to grasp, and generally viewed it as something on-the-outside and foreign.

Not surprisingly, when questioned about any technical details of Bitcoin, the non-user participants expressed that they only knew about Bitcoin on a surface level (if at all) and could not answer any questions dealing with it in detail. 
Indeed, our non-user participants frequently stated that they did not use bitcoins because they did not understand exactly how they work. 

On the other hand, we found that the user participants demonstrated an unusually low level of comprehension about the basic mechanics of the Bitcoin protocol.

Many of them only demonstrated a partial level of understanding when questioned about topics ranging from defining  commonly bandied-about terms to the mining process.

Four of our user participants clearly did not understand the true purpose of mining, claiming that it was only about solving the computational puzzle to obtain bitcoins instead of stating unequivocally that the process introduces new bitcoins into circulation as well as adding transactions to the blockchain.

They also could not speak in detail about the components involved (blocks, transactions, etc.).

Most of their understanding of the Bitcoin protocol did not line up with how it actually works and yet this did not stop any of our user participants from being able to buy, sell, and trade bitcoins for goods and services at various online locations. This stands in stark contrast to what non-user participants reported: in particular, that because they do not understand how Bitcoin works they therefore cannot use it. 

Not understanding how mediums of exchange function is not an uncommon phenomeneon among the general population. This is evidenced by questions we asked participants about how common electronic payment method (e.g.~credit card) works technically.

What is curious in this scenario is that non-user participants are claiming that a lack of knowledge is what prevents them from adopting Bitcoin but that is not the case for other payment methods. We speculate that this can be attributed to a few factors. Non-user participants do not see any important public ``endorsements'' of bitcoin as a payment method -- physical currency is the first method people are ever introduced to and is produced by governments, electronic methods like debit and credit are promoted by banking institutions as well as vendors. Furthermore, rising digital schemes such as Google Wallet and Apple Pay carry branding from trusted corporations that produce smartphone and Internet technology that pervades every aspect of daily life. Bitcoin runs very counter to these notions -- it is designed to be inherently anti-establishment by rejecting the supposed safety of a bank, there are no government controls on the currency that insure its value or safety, and it has no universal public spokesman to make arguments for it or to advocate on its behalf. Furthermore, there are not clear benefits to adopting Bitcoin to make payments; physical money can be carried around and transferred easily while debit and credit provide high convenience through direct debits to an account. Finally, what little branding it does have appears to be mostly in negative contexts among the general public as evidenced by our non-user participants who reported that they believed it was only used for black-market purchases, drug dealing, and money laundering. 

Our user participants had several misconceptions about the security and privacy controls of Bitcoin.

Many of our user participants expressed that Bitcoin inherently provides privacy since personal information is not leaked during transactions.  However, it has been demonstrated through some highly rigorous deanonymization work that the opposite is actually true: the base implementation of the protocol is, at best, weakly pseudonymous. The original argument by Nakomoto is that the creation of new addresses for every transaction should protect the users' privacy. Unfortunately, this is not true: BitIodine~\cite{spagnuolo}, wallet clustering~\cite{Meiklejohn:2013:FBC:2504730.2504747}, and network traffic~\cite{koshy} are a sample of the methods that can be used to track user payments on the blockchain; this is the central problem of having a publicly available ledger of accounts. There are some proposals to fix the anonymity problems through mixing networks~\cite{bonneau2} and onion routing; our users were mostly unaware of these.

A number of our participants were quick to bring up the ``51\% computing power'' argument as a reason that the blockchain could not be manipulated; generally, they appeared unconcerned with blockchain attacks. However, prior work has shown that colluding pools and dishonest miners can manipulate the public history with a much lower share of the total computing power~\cite{eyal, babaioff, eyal2014miner}.

There are no security protections built into Bitcoin to prevent fraudulent or mistaken transactions. Credits and debits to and from banking accounts can be halted or reversed with a call to the bank; with Bitcoin, there are no such avenues. As such, if money is sent to the wrong address then there is no way to reverse the transactions. There do exist some Bitcoin escrow services to mitigate this issue but none of our user participants mentioned it to us -- they were either not aware of them, did not trust them either, or simply preferred that the standard Bitcoin protocol allow for reversing transactions without the use of external services. Furthermore, user participants felt that Bitcoin was overall secure despite repeated examples of bitcoin thefts (Mt.~Gox, Bitstamp, etc.).  However, Bitcoin is more privacy conscious than other payment methods because there is no third party that can store personal information at an intermediate step between transactions as in the case of current electronic methods.

User participants felt strongly that they would prefer insurance for their bitcoins in the case of theft or fraud. Moreover, they actually said that they preferred if governments provided this as a service and claimed it would improve public trust in the currency. When followed up with questions about whether or not governments should regulate Bitcoin markets, they were quick to disparage government overreach and interference. This is a clear example of cognitive dissonance at play; allowing the government to insure deposits would give them a measure of control over users and exchanges, at which point they could then introduce side regulations by using the insurance as a cudgel in instances when they demand compliance.

We asked our participants about what features an ideal payment system should have in order for it to appeal to them. They advocated strongly for having fast transactions, especially when dealing with international transfers; many participants expressed discomfort with the wait times associated with using traditional wiring methods to transfer money between accounts and overseas. People said that they preferred to have multiple modalities for using the money -- both virtual and physical representations were desired as a matter of habit and convenience. Participants also were very concerned with financial fraud and privacy.

There was a particular amount of grumbling with respect to transaction fees; most participants prefer they be zero or as close to zero as possible as the status quo currently does not satisfy them. Bitcoin satisfies some of these properties that participants 
advocated for: fast transactions and low fees.

Bitcoin faces mainstream adoption problems because it does not appear to fit, based on descriptions of other payment methods by our participants, the needs of the general population. Debit, credit, and cash have all carved niches of use in day-to-day life: debit is a convenient method that draws directly from a banking account; credit can be used in instances where money is not immediately available; cash is accepted everywhere and can be transferred between people quickly. This is not to say that Bitcoin does not have superior use cases, it is just that these cases do not scale well to life in the United States: micropayments, intercontential money transfers, and person-to-person transactions. People mostly avoid these scenarios because they either do not need to do them or have been trained to work around them (Bitcoin is fairly new, after all). It is a new method of payment 
without clear benefits -- for example, a person may not commonly use credit cards but will hold onto it to build credit scores in the US; Bitcoin does not have such an allure.

Previous studies~\cite{mainwaring} on ideal aspects of digital payment schemes said that there are five aspects of a digital payment system required to appeal to users: (1) reduced complexity, (2) center around public use, (3) support money management without increasing burden or degrading user experience, (4) engage multiple senses, (5) and be fun to use~\cite{mainwaring}. The last two are aspects that are very particular to the study environment, Japan, which has many digital payment schemes that vie for control of the market. As evidenced by our participant responses, we can see where Bitcoin falls short. Bitcoin does not reduce complexity; it increases it by asking users to manage addresses and wallets. Bitcoin is centered around public use from its first design principles (user-first, no third parties). There are various clients to manage money; however, the introduction of additional  processes to use is clearly an increased burden on the user. Our questions do not generalize well to the last two categories; this is not really a problem, though, since the United States population generally does not appear to view money through that type of lens.

\paragraph{Limitations and Further Work}
As it is typical for interview studies, the availability and willingness of participants to participate in research can affect the results. Many of the user participants were very privacy conscious; this makes it difficult to do proper recruiting inside the Bitcoin community. We cannot be certain how this affects our findings. Our Bitcoin user sample does not fully represent all possible different agents in Bitcoin system: miners, investors, dogmatic believers, economists, computer scientists, and end users. However, our participants spanned more than seven different states across the US and vary widely in demographic terms. Thus, some further work would be warranted in characterizing the Bitcoin population. For example, how pervasive are the myths about Bitcoin advantages and disadvantages?

\paragraph{Conclusions}
We presented a study of users and non-users of Bitcoin using semi-structured interviews to discover more about their knowledge, attitudes, and opinions concerning Bitcoin. We found that a majority of our user participants held various misconceptions about how Bitcoin actually works. 
Our non-user participants bemoaned that they could not use Bitcoin since they do not know how it works; this is clearly not a barrier for the user participants so the real reason must lie elsewhere. User participants staunchly oppose government regulation but still desire them to insure any deposits. 

According to our participants, the major advantages of Bitcoin include fast transaction speeds, low transaction fees, security, and no third-party regulation. Perceived disadvantages of Bitcoin included that it is not widely accepted, its price keeps changing, and it has no mechanism for reversing payments. Interestingly, the properties that participants (e.g.~including non-users of Bitcoin) ascribed to an ideal payment system are ones that Bitcoin has.  We conclude that the Bitcoin user demographic pools are a highly fragmented collection of people that merit further examination while the non-user demographic pool could be surveyed in more detail about the barriers to entry necessary for them to use Bitcoin.

\bibliographystyle{plain}
\bibliography{cleaned_bitbib}

\end{document}